\documentclass [11pt]{article}
\usepackage{amsmath,amsthm,amsfonts,amscd,eucal,latexsym,amssymb}
\usepackage{epsfig}  
\def\cA{{\ca A}}

\def\cD{{\ca D}}

\def\cH{{\ca H}}
\def\cI{{\ca I}}

\def\cM{{\ca M}}

\def\cQ{{\ca Q}}

\def\cS{{\ca S}}

\def\cK{{\ca K}}

\def\bC{{\mathbb C}}           
 
\def\bI{{\mathbb I}}

\def\bR{{\mathbb R}}

\def\bS{{\mathbb S}}

\def\gA{{\mathfrak A}}       

\def\gF{{\mathfrak F}} 
 
\def\gH{{\mathfrak H}}

\def\gK{{\mathfrak K}} 
 
\def\gM{{\mathfrak M}}

\def\beq{\begin{eqnarray}}
\def\eeq{\end{eqnarray}}
\def\pa{\partial}
\def\at{\left(}               
\def\ag{\left\{}              

\def\ct{\right)}              
\def\cg{\right\}}             
\newcommand{\ca}[1]{{\cal #1}}         

\def\be{\beta}

\def\Ga{\Gamma}
\def\De{\Delta}
\def\La{\Lambda}

\def\Om{\Omega}

\newcommand{\nref}[1]{(\ref{#1})}
\newcommand{\media}[1]{\langle{#1}\rangle}

\newcounter{proposition}[section]
\newcounter{theorem}[section]
\newcounter{lemma}[section]
\newcounter{definition}[section]
\newcounter{corollary}[section]
\def\theproposition{\thesection.\arabic{proposition}}
\def\thetheorem{\thesection.\arabic{theorem}}
\def\thelemma{\thesection.\arabic{lemma}}
\def\thedefinition{\thesection.\arabic{definition}}
\def\thecorollary{\thesection.\arabic{corollary}}

\newcommand{\se}[1]{\section{#1}}

\def\vsp{\vspace{0.2cm}}
\def\vspp{\vspace{0.1cm}}

\def\sse #1 {\vsp\ifhmode{\par}\fi\refstepcounter{subsection}
  \noindent {\bf\thesubsection}. {\em #1}.\quad
  \addcontentsline{toc}{subsection}{\protect\numberline{\thesubsection} #1}%
  }

\def\ssb #1 {\vsp\ifhmode{\par}\fi\refstepcounter{subsection}
  \noindent {\bf\thesubsection.} {\bf #1.}\quad
  \addcontentsline{toc}{subsection}{\protect\numberline{\thesubsection} #1}%
  }

\def\ssa #1 {\ifhmode{\par}\fi\refstepcounter{subsection}
  \noindent {\bf\thesubsection.} {\bf #1.}\quad
  \addcontentsline{toc}{subsection}{\protect\numberline{\thesubsection} #1}%
  }

\def\proposizione #1 {\vsp\ifhmode{\par}\fi\refstepcounter{proposition}
  \vsp\ifhmode{\par}\fi\noindent {\bf Proposition \theproposition}. \quad {\em #1}}
\def\teorema #1 {\vsp\ifhmode{\par}\fi\refstepcounter{theorem}
  \vsp\ifhmode{\par}\fi\noindent {\bf Theorem \thetheorem}. \quad {\em #1}}
\def\lemma #1 {\vsp\ifhmode{\par}\fi\refstepcounter{lemma}
  \vsp\ifhmode{\par}\fi\noindent {\bf Lemma \thelemma}. \quad {\em #1}}
\def\definizione #1 {\ifhmode{\par}\fi\refstepcounter{definition}
  \vsp\ifhmode{\par}\fi\noindent {\bf Definition \thedefinition}. \quad {\em #1}}
\def\corollario #1 {\vsp\ifhmode{\par}\fi\refstepcounter{corollary}
  \vsp\ifhmode{\par}\fi\noindent {\bf Corollary \thecorollary}. \quad {\em #1}}

\def\proof #1 {\vspp\ifhmode{\par}\fi\noindent {\it Proof.} {#1} $\Box$\vsp\par}
\def\remark {\vsp\ifhmode{\par}\fi\noindent\noindent {\bf Remark:} 
}

\begin{document} 
 
\hfill{\sl Preprint UTM 705 - Oct 2006} 
\par 
\bigskip 
\par 
\rm 
 
 
\par 
\bigskip 
\LARGE 
\noindent 
{\bf On localization and position operators in\\ M\"obius-covariant theories.} 
\bigskip 
\par 
\rm 
\normalsize 
 
 
\large 
\noindent {\bf Nicola Pinamonti} \footnote{E-mail: pinamont@science.unitn.it} \\
 Department of Mathematics, Faculty of Science,
University of Trento, \&
 Istituto Nazionale di Alta Matematica ``F.Severi'',  unit\`a locale  di Trento \& Istituto
 Nazionale di Fisica Nucleare,  Gruppo Collegato di Trento, 
 via Sommarive 14, 
I-38050 Povo (TN), 
Italy\\\par 
\rm\large\large 
 
\rm\normalsize 

\rm\normalsize 
 
 
\par 
\bigskip 

\noindent 
\small 
{\bf Abstract}. 
Some years ago it was shown that, 
in some cases, a notion of locality 
can arise from 
the symmetry group of the theory \cite{BGL,Schroer97, FS02}, i.e. in an intrinsic way.
In particular, when M\"obius covariance is present, it is possible to associate some particular transformations 
with the Tomita-Takesaki modular operator and conjugation of a specific interval of an abstract circle.
In this context we propose a way to define an operator representing the coordinate conjugated to 
the modular transformations. 
Remarkably this coordinate turns out to be compatible with the abstract notion of locality.
Finally a concrete example concerning a quantum particle on a line is given.

\normalsize
\bigskip 

\se{Introduction}

In relativistic quantum theories, localization is usually introduced only at the level of second quantization, 
considering the smearing of field operators by means of real local functions.
In particular, the set of locally smeared free fields generates a suitable $*$-algebra of operators.
Moreover, operators smeared by functions with spatially separated domains have to commute.
We stress that local operators are characterized by the corresponding real local smearing wavefunctions, 
that generate a dense real subspace, {\it $\bR$-subspace} of the one-particle Hilbert space (corresponding to the free theory).
While the characterization of the one-particle Hilbert space (Wigner space \cite{Wigner}) uses as building blocks only the abstract symmetry enjoyed by the theory, the definition of local operators, as usually given, requires functions in the position representation. Thus the localization procedure, making use of functions in a particular representation, is not completely intrinsic.

For this reason, it was proposed some years ago by Brunetti, Guido and Longo \cite{BGL} and by Schroer \cite{Schroer97,FS02} 
 to introduce $\bR$-subspaces of local smearing wavefunctions in a more intrinsic way\footnote{See also \cite{Mund,MSY04,BL04,Schroer2006}.}.
 They suggested to interpret some suitable one-parameter subgroups of the symmetry group as 
the modular group introduced by Tomita and Takesaki \cite{Takesaki}. 
This modular structure is used to find local subspaces, hence called modular localization. 
The described procedure,  making use only of symmetry properties, is completely intrinsic and
no particular Hilbert space representation is required.
We remind the reader that there is another localization scheme called Newton-Wigner (NW) localization \cite{sig,fle,hal}, whose local objects are found by smearing with elements of proper subspaces of the one-particle Hilbert space. 
But since the NW localization is not fully compatible with relativistic theory, we shall not adopt this point of view.

Here we would like to consider quantum theories showing M\"obius covariance \cite{HL,BMT,GF,Longo,GLW,Carpi}, as usual this means that, on the Hilbert space of the theory $\cH$, acts an (anti)-unitary representation $\pi_\cM$ of the M\"obius group $\cM$. 
As Poincar\'e-invariant theories induce modular locality, 
also in $\cM$-invariant theories there is a suitable identification of the (proper) intervals of an abstract circle $\bS^1$ with particular subgroups of the representation. 
In fact, for every interval $I$, it is possible to select the following operators that represent particular M\"obius transformations: An anti-unitary operator $J_I$ and a one-parameter subgroup of unitary transformations $\De_I^{it}$.
In particular, $J_I$ and $\De_I$ play the role of Tomita-Takesaki modular conjugation and modular operator for the interval $I$. Then, reverting the usual point of view, local $\bR$-subspaces are defined as 
$$
\cK_I:=\{\psi\in\cH:J_I\De_I^{1/2}\psi:=\psi\}\;.
$$
We would like to stress that, in general, the elements of $\cK_I$ are not functions with domain contained in the interval 
$I$, moreover, there is no direct connection between intervals and operators, and the geometric interpretation of the action of $J_I$ and $\De_I$ on elements of the Hilbert space is not required.

The new aspect we are going to study is the presence of selfadjoint operators, {\it position operators}, representing suitable coordinates 
within the abstract circle $\bS^1$. 
Notice that Pauli's theorem \cite{Pauli} excludes the existence of a selfadjoint operator for a global coordinate 
that would be the canonical conjugate of the positive generator of rotations.
The same obstruction arises for the observable representing the time of arrival in quantum theories \cite{time} or for the relativistic coordinates of an event \cite{relevent}.
This kind of problem is usually addressed by generalizing the concept of observable that turns out to be described by means of positive operator-valued measures (POVM) \cite{Naimark,Ludwig,POVM}. 
Here we do not want to use this difficult method, instead we try  to define only {\it local coordinates}, choosing an interval $I$ and then defining the coordinate operator for the quantum theory restricted to $I$.
Furthermore, it is known that the vacuum theory on the circle appears as a thermal theory when restricted to an interval with respect to modular dilations. 
We know that in thermal theories a selfadjoint operator representing time can be defined \cite{NRT}, 
therefore we expect that a selfadjoint operator for the coordinate conjugate to the modular dilations $\De_I^{it}$ also exists.

It is known that, in the Newton-Wigner localization picture, there are meaningful coordinate operators. 
The situation appears completely different if one adopts the modular localization. 
Then the study of the interplay of the coordinate operators with modular locality seems to be relevant.
Since local objects are described by a dense $\bR$-subspace of the one-particle Hilbert space and not by proper complex subspaces,
we do not expect to establish the compatibility of coordinate operators $T$ with locality from their spectral properties.
Furthermore, we shall exhibit an interplay between modular locality and the range of the expectation values of $T$ that turns out to  be described by $a$ and $b$, when $T$ is evaluated in $\cK_{[a,b]}$ contained in $\cK_{I}$.
Hence the intrinsic locality does not only have an abstract meaning, but it turns out to describe the range of the expectation values of the coordinate operator $T$.
In this way it acquires a suitable physical meaning, even if it does not describe a Newton-Wigner coordinate.
Remarkably, $T$ turns out to be defined by a particular combination of the selfadjoint operators representing the generators of the group $PSL(2,\bR)$ contained in the M\"obius symmetry group.
In particular, for the coordinate within the semicircle $I_1$, and for a particular set $H,D,C$ of generators of $PSL(2,\bR)$ given below,
$T$ is defined as $T=(\log(H^{-1/2}CH^{-1/2}))/2$ while $D$ generates modular dilations $\De_{I_1}^{it}$.

This paper is organized as follows. First, some known results are presented in a useful way for the development of the subject. Then in the third section the most important results are presented.
The fourth section contains the discussion of a concrete example where locality arises by construction. 
It is shown that the proposed coordinate turns out to be compatible with the intrinsic modular locality. 
Then some final comments are presented.
In the appendix two kinds of irreducible representations of the covering group of $SL(2,\bR)$ on $L^2(\bR^+,dE)$ are discussed.

\se{Preliminary considerations}

\ssb{Interplay between M\"obius group, $PSL(2,\bR)$, modular operator and modular conjugation}
The M\"obius group is build up by conformal transformations of $\bC$ leaving the circle $\bS^1$ (modulus-one elements of $\bC$) invariant, it is generated by the group $PSL(2,\bR)$ and by an involution.
We recall that $PSL(2,\bR)$ is the left coset $SL(2,\bR)/\{+\bI,-\bI\}$, where $SL(2,\bR)$ is built up by real 
two-dimensional square matrices $g$ with determinant one satisfying $g\beta g^t=\beta$. $\beta$ is  the anti-diagonal matrix whose anti-diagonal elements are $(1,-1)$ (from top to bottom).
As already said, the circle $\bS^1$ can be viewed as the modulus-one elements of $\bC$, but also as the compactified line $\bR\cup\{\infty\}$, and the two pictures are connected by the Caylay transform $C: z\mapsto -i(z+1)(z-1)^{-1}$ that maps the elements of the circle to the compactified line.
The action of $PSL(2,\bR)$ on $\bS^1$ derives from the action of $PSL(2,\bR)$ on $\bR\cup\{\infty\}$:
\beq\label{azionesl2r}
x \to \frac{ax+b}{cx+d}, \qquad \begin{pmatrix} a & b \\ c & d \end{pmatrix} \in PSL(2,\bR).
\eeq
$g\in PSL(2,\bR)$ can be decomposed in the following way (Iwasawa decomposition) :
\beq\label{decomposizione}
g:=T(x) \La(y) P(z) ,\qquad x,y,z \in \bR,
\eeq
where
$$
T(x):=e^{xh}:= \begin{pmatrix} 1 & t \\ 0 & 1 \end{pmatrix}, \qquad
\La(y):=e^{yd}:= \begin{pmatrix} y & 0 \\ 0 & y^{-1} \end{pmatrix}, \qquad
P(z):=e^{zc}:= \begin{pmatrix} 1 & 0 \\ -z & 1 \end{pmatrix}.
$$
$h,c,d$ form a basis of the Lie algebra $sl(2,\bR)$, and they satisfy the following commutation relations:
\beq\label{algcomrel}
[h,d]=h,\qquad [c,d]=-c,\qquad [c,h]=2d\;.
\eeq
The other element generating M\"obius covariance is the reflection $r$ that maps $x$ to $-x$ in $\bR\cup\{\infty\}$.

\vsp
\noindent
In the subsequent part of the paper we shall use the representation of $\bS^1$ as the compactified line. 
Furthermore, we shall be interested in the set $\cI$ of proper intervals of $\bS^1$. 
Recall that an interval $I$ is {\it proper} if the interior of $I$ and is complement $I'$ are not empty.  
Every element $I$ of $\cI$ is denoted by $[a,b]$ where $a$ and $b$ are the 
initial and end point in $\bR\cup\{\infty\}$. The semicircle will be denoted by $I_1$, the particular set $[0,+\infty]$.
The action of $\La(x)$ is closed in $I_1$, while $r$ maps $I_1$ to its complement $I_1'$ in $\bS^1$. 

\proposizione{For every proper interval $I$ there exists $g\in PSL(2,\bR)$ such that $I=gI_1$. 
Moreover, the Iwasawa decomposition \nref{decomposizione}, 
can be generalized to every interval $I$ by $T_I(a)= g T (a) g^{-1}$, $P_I(a)= g P(a) g^{-1}$ and  $\La_I(a)= g \La(a) g^{-1}$. 
Furthermore, $r_I=g r g^{-1}$ maps $I$ to $I'$}

\vsp\noindent
The following properties, that can be checked by \nref{azionesl2r} and by the preceding proposition, hold for the general interval $I$: 
\begin{itemize}
\item[{\bf (A)}]
{\bf Reflection covariance:} $r_I$ maps $I$ to $I'$, and $r_{gI}=g r_I g^{-1}$ for every 
$g\in PSL(2,\bR)$.

\item[{\bf (B)}] {\bf  $\Lambda$ covariance:} The action of $\La_I$ is closed in $I$, 
and $\La_{gI}(t)=g\;\La_I(t)\;g^{-1}$ for $g$ in the triangular subgroup.

\item[{\bf (C)}] {\bf Positive inclusions:} We have two types of positive inclusions. 
In fact the action of $T_I(t)$ is closed in $I$ for positive $t$ and satisfies the following relation: 
$$
\La_I(b)T_I(t)\La_I(-b)=T_I(e^{2\pi b}t);
$$
while the action of $P_I(t)$ is closed in $I$ for negative $t$ 
and that satisfies the following relation: 
$$
\La_I(b)P_I(t)\La(-b)=P_I(e^{-2\pi b}t).
$$
\end{itemize}

\noindent
As a final remark we would like to recall that there is another particular one parameter subgroup of transformations $R(\theta)$ generated by $(h+c)/2$ whose action on a point is that of rotating the circle $\bS^1$ by an angle $\theta$. 
In particular, $R(\pi)$ maps the interval $I_1$ to its complement $I_1'$, $x$ to $-1/x$, and
$$
R(\pi) h R(-\pi)= c\;,\qquad  R(\pi) d R(-\pi)= -d\;,\qquad R(\pi) c R(-\pi)= h\;.
$$

\ssb{Quantization and locality: Weyl algebras, von Neumann algebras and the conformal net}
Up to now we only have described an abstract connection between a representation of the M\"obius group and intervals of a circle.
Now we would like to analyze its interplay with quantum theory. Hence we consider a proper positive-energy (anti)-unitary representation $U$ of the M\"obius group on  the one-particle separable Hilbert space $\cH$ of the quantum system treated.
In particular, $U$ contains a unitary positive-energy representation of $PSL(2,\bR)\subset\cM$.
Then, passing to the representation of the corresponding $sl(2,\bR)$ algebra, 
$h,d,c$ \nref{algcomrel} turn out to be represented by the selfadjoint operators 
$H$, $C$ and $D$, enjoying the following commutation relations:
\beq\label{sl2r}
[H,D]=iH,\qquad [C,D]=-iC,\qquad [H,C]=2iD.
\eeq
We will denote the domains of selfadjointness of $H$, $D$ and $C$
by $\cD(H), \cD(D), \cD(C)$ respectively. 
Furthermore, since $U$ contains a unitary representation of $PSL(2,\bR)$, there exists a common dense set of analytic vectors for $H$, $D$ and $C$. Thus the analytic vectors are contained in $\cD:=\cD(H)\cap\cD(D)\cap\cD(C)$ which turns out to be a dense set in $\cH$. Explicit representations of $H,D,C$ in the case of irreducible representations can be found in appendix 
\ref{aa}.
Notice that $H$ and $C$ are positive operators, while the spectrum of $D$ covers all of $\bR$.
To implement the properties (a), (b) and (c) described above we need to build up a local theory, in other words we would like to find real subspaces of $\cH$ representing local objects.
The first step is to define the Tomita operators associated with intervals $I$ in terms of the unitary representation of dilations $\La_I(t) \subset PSL(2,\bR)$ by means of $\De_I^{it}$, and the anti-unitary representation of reflections $r_I$ by $J_I$. Then we shall define  $S_I$ as
$$
S_I:=J_I\De_I^{1/2}\,.
$$
By means of $S_I$ we can define the real subspace of ``local functions'',
\beq\label{ki}
\cK_I:=\{ \psi \in \cH :S_I\psi=\psi \}\, 
\eeq
(considering $I_1$, $J_{I_1}$ maps $\cK_{I_1}$ in its complement and $\De_{I_1}^{it/(2\pi)}:=e^{-itD}$).

\proposizione{Let $\sigma$ be the imaginary part of the scalar product of $\cH$. 
Then, for every interval $I$,
the pair $(\cK_I,\sigma)$ forms a symplectic structure over $\cH$.}\\ 

\noindent
Notice that $\cH$, the target space of the unitary representation, has to be interpreted as the one-particle Hilbert space.

\vsp
\noindent
{\bf Weyl algebra.}
The concrete Weyl quantum fields $\hat{W}(\psi):=e^{i\hat{\varphi}(\psi)}$ on the standard Fock space $\gF(\cH)$ generate a unitary representation $\pi$ of the Weyl algebra $\gA(I)$ associated with the pair $(\cK_I,\sigma_I)$, called vacuum representation.
In fact it is nothing else but the GNS representation related to the vacuum state $\Om$.
$\hat{\varphi}(\psi)$ is the symplectically smeared field defined as $\hat{\varphi}(\psi):=ia(\psi)-ia^\dagger(\psi)$, $\psi\in\cK_I$, $a(\psi)$ and $a^\dagger(\psi)$ being creation and annihilation operator of the state $\psi$.
In this case locality arises by construction; in fact:

\proposizione{(Locality) Due to the symplectic form $\sigma$ a locality principle holds:
If $I_a$ and $I_b$  are disjoint subintervals of $I$, in $\gA(I)$ we have  
$$
[\hat{W}(\psi),\hat{W}(\psi')]=0 ,\qquad \text{if}\qquad  \psi \in \cK_{I_a}, \;  \psi' \in \cK_{I_b}.
$$
}

\noindent
{\bf Von Neumann algebra.}
Then the von Neumann algebra $\gM(I)$ associated with the interval $I$ over $\gH={\gF(\cH)}$ can be defined as
$$
\gM(I):=\{ W(\psi):\psi \in \cK_I \}''.
$$
Notice that the group representations are translated in a simple way to the von Neumann algebra through the action on $\psi\in\cK_I$.

\vsp
\noindent
{\bf Conformal Net}
Finally we would like to recall that the set $\cA$ of all local conformal algebras $\gM(I)$ for every interval $I$ defined above forms a particular conformal net. 
This can directly be seen by verifying the general requirements that characterize the conformal net of von Neumann algebras: Isotony, Locality, Conformal invariance, Positivity of the energy and Existence of the vacuum\footnote{More details can be found in \cite{BMT,GF,Longo,Carpi} and reference therein.}.
Notice that the following properties are easily satisfied by the elements of $\cA$:

\begin{itemize}
\item[{\bf (A)}] {\bf Isotony.} If $I_1\subset I_2$ then $\gM(I_1)\subset \gM(I_2)$.

\item[{\bf (B)}]  {\bf Locality.} If $I_1$ and $I_2$ are disjoint proper intervals, then  $\gM(I_1)\subset\gM(I_2)'$.

\item[{\bf (C)}] {\bf Conformal invariance.} There exists a strongly continuous unitary representation $U$ of $PSL(2, \bR)$ on $\gH$ such that $U(g)\gM(I)U(g)^*=\gM(gI)$, $g\in PSL(2,\bR)$.

\item[{\bf (D)}] {\bf Positivity of the energy.} If $R(\theta)$ denotes the rotation by an angle $\theta$ on $\bS^1$, the generator of the rotation subgroup $U(R(\theta))$ (also called conformal Hamiltonian) is positive.

\item[{\bf (E)}] {\bf Existence of the vacuum.} There exists a unit vector $\Om\in\gH$ (vacuum vector) which is invariant under the unitary representation of $PSL(2,\bR)$ and cyclic for every $\gM(I)$ where $I$ is a proper interval.

\end{itemize}

\noindent
A consequence of the presence of a conformal net is the Reeh-Schlieder property \cite{BMT,GF,Longo,Carpi}. 
In particular, it says that the the vacuum vector $\Om$ is cyclic and separating for any local algebra $\gM(I)$.
Hence, for every $\gM(I)$, $\Om$  is a KMS state with respect to the modular group $\De_I^{it}$. 
In particular, for the case of the von Neumann algebra built over half of the circle $\gM_{I_1}$, $\Om$ can be interpreted as a thermal state at temperature $1/(2\pi)$ with respect to the unitary transformation generated by $D$.
$D$ can be interpreted as the energy of the particular system.

\vsp
\noindent
As a final comment we notice that $\cK_I\subset\cH\subset{\gH}$ and that the Hilbert subspace $\cH$ does not contain $\Om$.
Furthermore $\cK_I:=P\gM(I)\Om$, where $P$ projects elements of $\gH$ in $\cH$.
In the next part of the paper we shall use only the fact that if $\psi=A\Om$ with $A\in \gM(I)$,
then $J\De_I^{1/2}\psi=\psi$, and the fact that $H$ and $C$ have continuous spectrum on $\cH$.
Hence the result presented below for  $\cK_I\subset\cH$ can be generalized to $(\gM(I)\Om\cap \{\Om\}^\perp)\subset \gH$.  

\se{Modular coordinate}

The aim of the present section is to find operators to be interpreted as coordinates of intervals in the case of theories invariant under the M\"obius group. We notice immediately that a global coordinate of $\bS^1$ cannot be represented by a selfadjoint operator; 
in fact such an operator should be conjugate to the generator of rotations which exists in the algebra of observables 
and is bounded from below. 
In this case Pauli's theorem \cite{Pauli} excludes the existence of a conjugate selfadjoint operator.
This problem can be circumvented generalizing the concept of observables by means of POVM \cite{Naimark,Ludwig,POVM}. 
Here we do not want to address the problem in this way. Instead we are looking for coordinates of intervals whose corresponding translations are described by the modular group $\De_I^{it}$ and whose corresponding generator 
is not bounded from below. 
From now, without loss of generality, we fix a specific interval $I_1$ which corresponds to the semicircle or equivalently to half of the line in the projective representation. 
The associated modular operator is  $\De:=e^{-2\pi D}$, where $D$ is the generator of dilations described above.
We are looking for a generator $T$ of a group of unitary transformations $W(a)$ that forms a Weyl-Heisenberg group with $\De^{it/(2\pi)}$:
\beq\label{ccr}
\De^{it/(2\pi)} W(a)=e^{iat}W(a)\De^{it/(2\pi)}.
\eeq
Notice that the operator we sought after should have a continuous spectrum that coincides with $\bR$. Then an important step in our work is to check if this operator when evaluated on wave-functions local in $I=[a,b]\subset I_1$ gives results in the real interval $(\log a,\log b)$.
If such an operator exists it could be interpreted as the modular coordinate for the interval $I_1$.

\vsp
\noindent
Let us start with the following proposition, that will also be useful later:

\proposizione{\label{sqrthc} Suppose that on $\cH$ there acts a positive-energy unitary representation of the covering group of $SL(2,\bR)$, generated by $H,C,D$, which are selfadjoint operators satisfying \nref{sl2r} on $\cD:=\cD(H)\cap\cD(C)\cap\cD(D)$.
Consider the selfadjoint operator $H^{-1/2}$ with domain $\cD(H^{-1/2})$ defined via the spectral theorem. Then $\cD\subset \cD(H^{-1/2})$. Let $\psi\in\cD$, then 
$$
\|H^{-1/2}\psi\|^2\leq \at k-\frac{1}{2}\ct^{-2}{(\psi,C\psi)}\;,
$$
where $k\geq 1$. $k$ is the lowest eigenvelues of $(H+C)/2$.}
\proof{Decompose the Hilbert space $\cH=\bigoplus_i \cH_i$, where the subspaces $\cH_i$ are $PSL(2,\bR)$-irreducible.
The irreducible representation on $\cH_i$ is labelled by $k_i$; the lowest eigenvalue of $(H+C)/2$ on $\cH_i$.
Then $k$ denotes the smallest $k_i$. 
Since we are considering positive-energy representation of $PSL(2,\bR)$, every $k_i\geq 1$, hence also $k\geq 1$. 
The following relation holds
$$
\|DH^{-1/2}\psi \|^2+\at k-\frac{1}{2}\ct^2\|H^{-1/2}\psi\|^2\leq {(\psi,C\psi)},
$$
proving the statement.}

\remark
In a similar way it can be proved that $\cD$ is also contained in the domain $\cD(C^{-1/2})$ of selfadjointness of $C^{-1/2}$.

\vsp
\noindent
The positive inclusions, generated by $h$ and $c$, introduced at group level are also present as unitary transformation of the Hilbert space.

\proposizione{\label{posinc} Given a positive-energy unitary representation of $PSL(2,\bR)$ on $\cH$ generated by $H$, $C$ and $D$ satisfying \nref{sl2r}, let $U_h(a)$ and $U_c(b)$ be the two one-parameter unitary subgroups generated by $H$ and $C$. Then:\\
\noindent
(a) $U_h(a)\cK_{I_1} \subset \cK_{I_1}$ for every strictly positive $a$,  and
$$
\De^{it} U_h(a) \De^{-it} = U_h(e^{-2\pi t} a) \forall t \in \bR\;, \qquad J U_h(a) J= U_h(a)^* ;
$$
(b) $U_c(a)\cK_{I_1} \subset \cK_{I_1}$ for every strictly negative $a$,  and 
 $$
\De^{it} U_c(a) \De^{-it} = U_c(e^{2\pi t} a) \forall t \in \bR\;, \qquad J U_c(a) J= U_c(a)^* .
$$
}

\noindent
From the positive inclusion properties enjoyed by the representation of $PSL(2,\bR)$, it is possible to build up a one-parameter group of unitary transformations forming a Weyl-Heisenberg group with the modular transformation $\De^{it}$. 
In fact the following theorem holds.

\teorema{ \label{positiveinclusion} Under the hypothesis of proposition \ref{posinc}, 
each of the two selfadjoint operators defined by $T_h = log H$ and $T_c = log C$ generates a one-parameter unitary group satisfying the Weyl commutation relations with the dilation one-parameter unitary group $V(t):=\De^{it/(2\pi)}$.}
\proof{%
$V(t):=\De^{it/(2\pi)}$ acts as a one-parameter group of unitary transformations on $\cH$. 
To find the other group $W(a)$ we study the properties of $H$, the generator of $U_h(a)$.
From proposition \ref{posinc}, $\De^{it} H \De^{-it}= e^{-2\pi t} H$, it follows for the positive spectrum $\sigma(H)$ of $H$ that $\sigma(H)=e^{-2\pi t}\sigma(H)$. 
Since the representation we are considering is non trivial, the spectrum on $\cH$ has only an orbit $(0,+\infty)$.
The spectral decomposition of $H$ on $\cH$ is $H:=\int_{\bR^+} \lambda dE(\lambda)\;; $
then the operator 
$$
T_h:=\log(H)=\int_{\bR^+} \log(\lambda) dE(\lambda)\; \qquad \text{on}\qquad \cD(T_h):=\ag\psi: \int_{\bR^+} |\log(\lambda)|^2 (\psi,dE(\lambda)\psi)\cg\;
$$ 
is selfadjoint, and $\cD(T_h)$ is dense in $\cH$.
Let $W(a):=e^{iaT_h}$, from proposition \ref{posinc} we infer that $V(t)W(a):=e^{ita}W(a)V(t)$ is a projective unitary representation of the Weyl-Heisenberg group. The same holds also for $T_c:=\log C$.}
\noindent
Then because of proposition \ref{sqrthc} and of the spectral properties one derives the following

\corollario{Under the hypotheses of theorem \ref{positiveinclusion}, $\cD\subset\cD{(T_h)}$ and $\cD\subset\cD{(T_c)}$.}

\vsp
\noindent
Within the group representation there are two possible operators enjoying the canonical commutation relation with $D$. 
They offered plenty of candidates for describing a coordinate of $I_1$. 
In fact every
$T_x:=x \log(C)-(1-x) \log(H) + f(D)$ with $0\leq x\leq 1$, where $f(D)$ is an almost general function, if selfadjoint on a dense domain $\cD(T_x)$, together with $D$ generates a Weyl-Heisenberg group.
Thus the commutation relations alone cannot yield a particular meaningful coordinate operator.
In the next subsection we shall study a criterion to choose a preferred one.

\ssb{Interplay with locality} 
We have seen that for every interval there are several possible different position operators $T_x$. 
A good criterion to choose a preferred one is to study the compatibility with locality in the sense of the following definition.
\definizione{A selfadjoint operator $T_x$ is said to be compatible with locality if, considering $I=[a,b]\subset I_1$, its expectation values on $\cK_I$ lie between $\log(a)$ and $\log(b)$.}\\
\noindent 
The aim of the present sub-section is then to find an operator compatible with locality that shows canonical commutation relation with the generator of dilation.

\noindent
Since, for systems having M\"obius covariance, locality arises intrinsically from the group properties and since also the coordinate operators derive from the group generators itself, we would like our choice to be completely determined by the (anti)-unitary representation of the M\"obius group.
%
%
\noindent
We start with some preliminary propositions. From now on $\cK:=\cK_{I_1}$ defined as in \nref{ki}. 

\proposizione{ \label{Dpositive} Let $D$ be the generator of dilations of a unitary representation of $SL(2,\bR)$ and $\cK$ as described above. Then $D$ is positive on  $\cK\cap\cD(D)$.}
\proof{ Since $\psi\in\cK$, $J\De^{1/2}\psi=\psi$, then $\|D\psi\|=\|D\De^{1/2}\psi\|$, but $\psi\in\cD(D)$ then also $\De^{1/2}\psi\in \cD(D)$. 
Be $\alpha$ in the strip $\cS:=\{z\in\bC : 0<\Re(z)<1\}$, by spectral calculus 
$$\|D\De^{\alpha/2}\psi\|^2\leq \|D\psi\|^2+\|D\De^{1/2}\psi\|^2,$$
which means that $\De^{\alpha/2}\psi\in\cD(D)$ with $\alpha<1$.
Consider the function $F(\alpha):=(\psi,D\De^{\alpha}\psi)$, it is analytic in the strip $\cS$ and in particular real and smooth for $\alpha$ real and $0<\alpha<1$.
Notice that $|F(0)|=|(\psi,D\psi)|$ is finite by hypothesis and that 
$$
F(0)=(J\De^{1/2}\psi,D\; J\De^{1/2}\psi)=( - D \De^{1/2}\psi,\De^{1/2}\psi)= -F(1).
$$ 
Moreover $\frac{d F(\alpha)}{d \alpha}=-2\pi(\De^{\alpha/2}\psi,D^2\De^{\alpha/2}\psi)$ is negative in the interval $(0,1)$.
We can conclude that $F(0)$ is positive, hence the statement of the proposition.}
\remark The above proposition holds also for local wave-functions $\cK$ modified by a phase, namely for $e^{i\alpha}\cK$.
The proof can be expended also in $\gF(\cH)$ to every $\psi\in\cD(D)$ such that $\psi=A\Om$, where $A\in\gM(I_1)$.
An alternative proof of proposition {\ref{Dpositive}} can already be found in the literature (Proposition 1.14 of \cite{GL}).

\proposizione{\label{disugualianze}Let $\psi$ be in the domain of $H$, $D$  and $C$, the generators of a unitary representation of $PSL(2,\bR)$ satisfying \nref{sl2r}, if $\psi$ is in $\cK_{I}$, where the interval $I=[a,b]$ is  properly contained in $I_1$, with $0<a<b<+\infty$, the subsequent inequalities hold
$$
a^2(\psi,H\psi) \leq (\psi,C\psi)\leq b^2(\psi,H\psi)\;.
$$
}
\proof{The unitary transformation $U:=\exp{(iaH)}$ maps the semicircle $I_1$ to $I_a:=[a,\infty)$ and  $\cK$ is mapped to the corresponding $\cK_{I_a}$ by means of $U$.
Moreover $\cK_I\subset \cK_{I_a}$, so that for $\psi\in\cK_{I}$, $U^\dagger \psi \in \cK$. Under this transformation 
$$U^\dagger CU:=C+2aD+a^2 H, \qquad U^\dagger HU=H$$
(see \cite{MP} for details). 
Since by proposition \ref{Dpositive} $(U^\dagger \psi,D\,U^\dagger \psi)\geq 0$, we have 
$a^2(\psi,H\psi)\leq (\psi,C\psi) $. 
The other inequality can be proved rotating $I_1$ by $\pi$, namely using the unitary transformation $\exp{i\pi(H+C)/2}$ that maps $I_1$ to its complement $I_1'$, $D$ to $-D$, $C$ to $H$ and vice versa.}

\noindent The preceding theorem can be generalized in straightforward way along the following lines. First of all notice that the proof can be extended also in the Fock space $\gF(\cH)$ to every $\psi\in\cD(D)\cap\cD(H)\cap\cD(C)$ 
such that $\psi=A\Om$, where $A\in\gM(I_1)$.
Thus the inequalities established in Theorem \ref{disugualianze} also hold in the Fock space. Since the only important property in the preceding proof is the fact that $D$ is positive on $\psi\in\cK$, a second generalization arises recalling that $D$ is also positive on wave-functions multiplied by a constant phase: $e^{i\alpha}\psi$. That would be used below to prove the most important theorem.

\noindent
The last comment we give concerns the fact that there is no need for a normalized wave-function in order to establish proposition \ref{disugualianze}. 
Furthermore, the logarithm is a monotonic function and also an operator monotonic function so that proposition \ref{disugualianze} suggests that 
$$\log(a) \leq (\log\media{C}_\psi-\log\media{ H}_\psi)/2\leq \log(b)\;,$$
where $\media{C}_\psi=(\psi,C\psi)$.
Then we would like to see if it is possible to find an operator $T$ arising from $H$ and $C$ that commutes with $D$ and 
satisfies $\log(a)\leq \media{T}_\psi\leq \log(b)$ for $\psi\in\cK_{[a,b]}$.

\noindent Next we shall see that $T$ is not simply $(\log C-\log H)/2 $ on a suitable common domain, but arises 
from another representation of the covering group of $PSL(2,\bR)$ that uses $H$, $D$ and $C$ as building blocks.

\definizione{A selfadjoint operator $O$ is said to be local in a $\bR$-linear subspace $\cK$ of the Hilbert space if $O\cK\subset\cK$.}

\proposizione{$H^2$, $C^2$, $D^2$ are local operator in $\cK_I\cap\cD$.}
\proof{Since $J\De^{1/2}H^2\psi=H^2\psi$, $H^2$ is local in $\cK_I\cap\cD$, and the same holds for $C^2$ and $D^2$.} 

\noindent
The preceding proposition does not hold for $H$, $D$ and $C$.

\proposizione{\label{repinterplay}Let $U$ be a unitary, non trivial, representation of $PSL(2,\bR)$ on the separable Hilbert space $\cH$, generated by the operators $H, C, D$ satisfying \nref{sl2r} which are essentially selfadjoint on a common domain $\cD$. 
The following operators 
$$
\widetilde{H}:=\frac{H^2}{2},\qquad \widetilde{D}:=\frac{D}{2},\qquad \widetilde{C}:=\frac{H^{-1/2}CH^{-1/2}}{2}\;,
$$ 
are essentially selfadjoint on a domain $\widetilde{\cD}$ of $\cH$.
Moreover, they satisfy the commutation relation \nref{sl2r} of  $sl(2,\bR)$. They generate a positive-energy unitary representation $\widetilde{U}$ of the covering group of $SL(2,\bR)$ on $\cH$. }

\proof{Since the Hilbert space $\cH$ is separable, the representation $U$ on  $\cH$ can be decomposed into a direct sum of irreducible representations. 
We can consider a component $U_i$ of this decomposition as an irreducible representation on a Hilbert space $\cH_i\subset\cH$. 
Let $k_i$ be the highest weight of the representation $U_i$. 
Since $k_i\geq 1$, every $\cH_i$ is isomorphic to $L^2(\bR^+, dE)$, where $H,D,C$ take on the usual form described in the appendix \nref{gensl2r1} with $k=k_i$, while $\widetilde{H}, \widetilde{D},\widetilde{C}$ take on the other form \nref {gensl2r2}. 
Consider $\widetilde{\cD}$ as the sum of analytic vectors. 
Then, by some result of Nelson \cite{nelson}, $\widetilde{H}, \widetilde{D},\widetilde{C}$ turn out to be the generators of a unitary representation of $\widetilde{SL(2,\bR)}$, the covering group $SL(2,\bR)$ on $\cH$.}
\noindent 
\remark
{\bf (a)} We recall that the irreducible representations of the covering group of $SL(2,\bR)$ are labelled by the lowest eigenvalue of the generator of rotations $(H+C)/2$, and the representation is of positive energy if $k\geq 1/2$.
Let $k$ and $\widetilde{k}$ be the lowest eigenvalues of the generators of rotations $(H+C)/2$ and $(\widetilde{H}+\widetilde{C})/2$ respectively, $\widetilde{k}=k/2+1/4$.
Since $H,C,D$ generate a unitary representation of $PSL(2,\bR)$, $k\geq 1$ and $\widetilde{k}\geq 3/4$. 
Hence $\widetilde{H},\widetilde{C},\widetilde{D}$ generate a positive-energy representation of the covering group of $SL(2,\bR)$.

\noindent {\bf (b)} 
Let us analyze the relation between the domain of the representations generated by the selfadjoint operators $H$, $C$ and $D$ and $\widetilde{H}$, $\widetilde{C}$ and $\widetilde{D}$ respectively defined on suitable domains $\cD$ and $\widetilde{\cD}$ on $\cH$ as in proposition \ref{repinterplay}.
As far as we know there is no clear connection between $\cD$ and $\widetilde{\cD}$, but something can be said about the domain of the quadratic forms associated with the operators considered.
We recall that having a selfadjoint operator $X$ defined on the domain $\cD(X)$, the associated quadratic form $(\cdot,X\cdot)$ can in principle be extended to a quadratic form $X(\cdot,\cdot)$ defined on a domain $\cQ(X)\times\cQ(X)$
larger then $\cD(X)\times\cD(X)$. 
Now, 
consider $\cQ\subset\cH$ formed by the elements $\psi$ of the Hilbert space such that the quadratic forms $H(\psi,\psi)$, $C(\psi,\psi)$ and $D(\psi,\psi)$, which extend to $(\cdot,H\cdot)$, $(\cdot,C\cdot)$ and $(\cdot,D\cdot)$, are defined and finite.
For this set, the following relation holds:

\proposizione{Let ${\cQ}$ be as described above, and let $\widetilde{\cD}$ be the domain of selfadjointness of the operators $\widetilde{H}$, $\widetilde{D}$ and $\widetilde{C}$ defined in Proposition \ref{repinterplay}. 
Then $\widetilde{\cD}\subset \cQ.$}
\proof{Consider $\psi \in \widetilde{\cD}$, then obviously $H(\psi,\psi)\leq(\psi,\widetilde{H}^{1/2}\psi)$ is well-defined and finite. 
The same is true of  $D(\psi,\psi)\leq(\psi,2\widetilde{D}\psi)$ which is also finite.
The only difficult part in this proof is to check that $C(\psi,\psi)$ is finite. 
To this end notice that
$$C(\psi,\psi)\leq\alpha\|\widetilde{C}\psi\|\,\|\widetilde{H}^{1/2}\psi\|+\beta\|\widetilde{D}\psi\|\,\|\widetilde{H}^{-1/2}\psi\|\;,$$
where $\alpha$ and $\beta$ are finite numbers. 
The right hand side of the inequality written above is clearly finite, hence $\psi\in\cQ$ and the statement is proved.}

\vsp
\noindent Now for the generator $\widetilde{C}$ of the new representation it is a surprise to 
have a proposition similar to proposition \ref{disugualianze}.
\proposizione{\label{disugualianze2} Let $\widetilde{C}$ be the selfadjoint operator defined on $\widetilde{\cD}$ as in Proposition \ref{repinterplay}, and let $\psi\in\cK_{I}\cap\widetilde{\cD}$, where $I=[a,b]\subset I_1$. Then 
$$
\frac{a^2}{2} \|\psi\|^2<(\psi, \widetilde{C}\psi)<\frac{b^2}{2}\|\psi\|^2\;.
$$}
\proof{$\widetilde{C}=H^{-1/2} C H^{-1/2}/2$, the first inequality can be proved by unitary transformation $\varphi:=U^\dagger\psi$ and $U^\dagger \widetilde{C} U$, where  $U:=e^{iaH}$. Notice that $\varphi \in \cK$ and that $U^\dagger \widetilde{C} U = H^{-1/2} (C+aD+a^2H) H^{-1/2}/2$.
Since $\varphi \in \cK\cap\widetilde{\cD}$ and $\widetilde{\cD}$ is contained in the domain of selfadjointness of $H^{-1/2}$ and $iH^{-1/2}\varphi\in\cK\cap\cD$, in a similar way as in proposition \ref{Dpositive} we get that
$$
(\varphi,H^{-1/2}DH^{-1/2}\varphi)
$$
is positive. Then $\frac{a^2}{2} \|\psi\|^2\leq(\psi, \widetilde{C}\psi) $ holds. The other inequality can be proved rotating $I_1$ by $\pi$.
}

\noindent Because of the previous discussion it appears clear that it is only necessary to pass to the logarithm of the operator $\widetilde{C}$ in order to define an operator describing the position compatible with locality. To this end consider the following theorem.

\teorema{\label{defT}Let $\widetilde{C}$ be the selfadjoint operator defined in proposition \ref{repinterplay}, and let $T$ be the operator 
$$
T:=\frac{1}{2}\log(2 \widetilde{C}).
$$
This operator is selfadjoint on a suitable domain $\cD(T)$ that contains $\widetilde{\cD}$. 
Moreover, it generates a unitary group of transformations, $W(a):=e^{ia T}$, which together with $V(t):=\De^{it/(2\pi)}$ generates a two-dimensional Weyl-Heisenberg group}
\proof{$\widetilde{C}$ is positive and selfadjoint on $\cD(\widetilde{C})$. 
Consider its spectral measure $P(\lambda)$, then 
$$
T:= \frac{1}{2} \int_{0}^\infty \log (2\lambda)\; dP(\lambda)
$$
is defined on 
$\cD(T):=\ag \psi, \frac{1}{2} \int_{0}^\infty |\log (2\lambda)|^2 d(\psi,P(\lambda)\;\psi) <\infty\cg.$
Since, by proposition \ref{sqrthc}, $\widetilde{\cD}\subset\cD({\widetilde{C}}^{-1/2})$, $\widetilde{\cD}$ is also contained in $\cD(T)$.
The proof of the second part of the theorem follows in a straightforward way by noticing that there is a representation of $\widetilde{SL(2,\bR)}$ generated by $\{\widetilde{H},\widetilde{D},\widetilde{C}\}$. 
Let $U(a):=e^{ia\widetilde{C}}$, then  $V(t)U(a)V(-t)=U(e^{-2t}a)$, and thus $W(a)$ and $V(t)$ generate a two-dimensional Weyl Heisenberg group.}

\teorema{\label{localvalues} Let $T$ be defined as in theorem \ref{defT}. 
Consider $\psi\in\cK_I\cap\widetilde{\cD}\subset\cK_{I_1}$, where $I$ is the interval $[a,b]\subset I_1$.
For the expectation values of $T$ in $\psi$ there holds the relation
\beq\label{diseqlog}
\log(a)\;\|\psi\|^2\leq (\psi,T\, \psi)\leq\log(b)\;\|\psi\|^2\;.
\eeq
}
\proof{Consider the function
$$
F(\alpha):=\frac{a^{-2\alpha}}{\|\psi\|^2}(\psi, (2\widetilde{C})^\alpha\psi)\;.
$$
Since $\widetilde{C}$ is positive and $\psi\in\widetilde{\cD}$, by spectral properties of $\widetilde{C}$, $F$ is a 
well-defined smooth real function for $-1\leq\alpha\leq 1$.
Since $\psi\in\cD(T)$ and $\log (2\widetilde{C})=2T$ on that domain, we have $\frac{dF}{d\alpha}(0)=-2\log(a)+(\psi,2T\psi)/\|\psi\|^2$.
As a second step notice that $F(0)=1$ while, by using proposition \ref{disugualianze2} on the state $\widetilde{C}^{-1/2}\psi$, $F(-1)\leq 1$.
Since $\frac{d^2F}{d\alpha^2}\geq 0$ in the interval $(-1,1)$ the first inequality in \nref{diseqlog} is proved.
The second can be proved in a similar way performing a rotation $U(R(\pi))$.
}

\noindent
The following corollary follows from the definition of $T$,  the action of  $g\in PSL(2,\bR)$ on $I$ and due to theorem 
\ref{localvalues}.
\corollario{Let $\psi\in\cK_I$ with  $\|\psi\|^2=1$. Then $(\psi,T\psi)$ transforms covariantly under unitary transformations of $PSL(2,\bR)$ that map $I=[x,y]$ in $I_i\subset I_1$. In other words, if $I_i=gI$, where $g=\begin{pmatrix}a&b\\c&d\end{pmatrix}$ and $U_g$ is the corresponding unitary transformation, then
$$
\log (x)\leq (\psi,T\psi)\leq \log (y)\qquad \text{implies} \qquad \log \at\frac{dx-b}{a-cx}\ct\leq (\psi,U_gTU_g^\dagger\psi)\leq \log \at\frac{dy-b}{a-cy}\ct\;.
$$
}

\remark Theorem \ref{localvalues} can be generalized to every element $\psi$ of $\gK:=(\gM(I_1)\Om\cap\{\Om\}^\perp)\subset\gH$.
Indeed, (1) on $\gK$ the spectrum of $\widetilde{C}$ is continuous, furthermore, $\De$ and $J$ are the modular operator and conjugation in $\gH:=\gF(\cH)$ for $\gM(I_1)$,
(2) $J\De^{1/2}\psi=\psi$ for every $\psi\in\gK$.
(1) and (2) are the only properties used to produce the result. 
This assures that $T$ defined as above is compatible with locality.
Hence, since $T$ enjoys canonical commutation relations with $D$, $T$ may be interpreted as a well-behaved coordinate associated with the interval $I_1$.

\vsp

\se{Physical situations}

In this section we introduce a concrete example, where M\"obius covariance arises and hence present a coordinate operator $T$ for the semicircle.
Without loss of generality, in this case, we assume that locality arises by construction and not intrinsically from the 
(anti)-unitary representation of 
the M\"obius group. We shall show that the real Hilbert subspace derived from intrinsic locality in the interval $I$ contains the local wave-function with support in $I$.

\ssb{Coordinate operator of a quantum particle on the half line}\label{opt}
We study the quantization of a one-dimensional particle associated with the so-called tachyon field.
This theory was proposed some years ago by Sewell \cite{Sewell}, restricting some Minkowskian free field theory on the Killing horizon. 
Sometimes in the literature this is related to lightfront holography \cite{Sch}.
Recently its relation with conformal theory was put forward \cite{GLRV,MP}. 
As to conformal theory on the circle, see for example \cite{BMT,GF,Longo,Carpi} and references therein. 
Consider the classical wave-functions $\psi$ in $\cS$: the set of smooth real functions vanishing at infinity along with with every derivative. 
Equip $\cS$ with the symplectic form 
$$
\sigma(\psi,\psi'):=\int_\bR \at \psi\pa_x \psi'-\psi'\pa_x\psi\ct\; dx\;,
$$
which is invariant under change of the coordinate $x$.
Once the coordinate $x$ is chosen, every wave-function can be decomposed into positive- and negative-frequency parts:
$\psi(x):=\psi_+(x)+\overline{\psi_+(x)}$, where the positive-frequency part is $\psi_+(x):=\int_0^\infty \frac{e^{-iEx}}{\sqrt{4\pi\, E}} \widetilde{\psi}_+(E) dE$.
The set of functions $\psi_+$ turns out to be (dense in) the Hilbert space 
$$
\cH_\bR\simeq L^2(\bR^+,dE)
$$ 
with scalar product $\langle\psi_+,\psi'_+\rangle:=-i\sigma(\overline{\psi}_+,\psi'_+)$.
As discussed in \cite{MP}, on $L^2(\bR^+,dE)$ there acts a faithful (anti)-unitary representation of the M\"obius group, in particular the one with lowest eigenvalues of $(H+C)/2$ equal to one; the explicit form of the corresponding generators is \nref{gensl2r1} in the appendix.
Furthermore, the action of $PSL(2,\bR)$ has a local meaning; in particular, the wave-function $\psi$ transforms according to
$$ 
U_g \psi(x):= \psi\at\frac{ax+b}{cx+d}\ct -\psi\at\frac{a}{c}\ct\;,\qquad g=\begin{pmatrix} a & b \\ c & d \end{pmatrix} \in PSL(2,\bR),  
$$
under the action of $g\in PSL(2,\bR)$, while $J$ is a reflection: $J\psi(x)=\psi(-x)$.

There are two ways to build real local subspaces of wave-functions, one that arises from geometrical considerations and the other from the group properties of $PSL(2,\bR)$.
In the former case,  the wave-functions in $\cS_I$ are defined as the elements of $\cS$ with support contained in $I$. In the latter case, we use the set $\cK_I$ defined in \nref{ki}. 
Let $\varphi\in\cS_I$, then $J_I\De_I^{1/2}\varphi_+=\varphi_+$. In a certain sense, by a little abuse of notation, 
we can say that for every $I$, $\cS_I\subset\cK_I$. 
The result found in the previous section holds true also for the elements of $\cS_I$:
Consider $\varphi\in\cS_I$, where the interval $I=[a,b]\subset\bR^+$,  we get 
$(\varphi_+,D\varphi_+)\geq0$ by proposition \ref{Dpositive} and $a^2 (\varphi_+,H\varphi_+)\leq (\varphi_+,C\varphi_+) \leq b^2 (\varphi_+,H\varphi_+)$ by proposition \ref{disugualianze}. 
It is furthermore possible to define the operator $T$ related to the half line coordinate defined as in \ref{defT} 
$$
T:=\frac{1}{2}\log(2\widetilde{C})\;,\qquad \text{where} \qquad \widetilde{C}:=-\frac{d^2}{dE^2}
$$
is a selfadjoint operator on $\cH_\bR$.
Finally, by theorem \ref{localvalues}, for every $\phi:=P\hat{W}(\psi)\Om$, where $P$ is a projector onto $\gK$, there holds  
$\log a \leq(\phi,T \phi)\leq \log b $ if $\|\phi\|=1$.

\noindent
Since $\widetilde{C}$ can be extended on $\gF(\cH_\bR)$ and has zero eigenvalues only on $\alpha\Om$, the operator $T:=\log(2\widetilde{C})/2 $ turns out to be selfadjoint on a domain $\cD$  dense in the sub-Hilbert space $\{\Om\}^\perp$.
Finally $\log (a) \|\phi\|^2\leq(\phi,T \phi)\leq \log (b)\|\phi\|^2 $ is valid for $\phi\in\gA(I)\Om \cap\{\Om\}^\perp$.

\noindent
We conclude that surprisingly $T$, which is defined {\it intrinsically} by means of the representation of $PSL(2,\bR)$, is self-adjoint on a suitable domain and defines a modular coordinate which is compatible with locality.
This is true even if local properties of $T$ are not manifest considering its action on a wave-function $\varphi\in\cS$.

\se{Final comments}

We have seen that, in case of M\"obius-covariant theories, an observable representing the coordinate of a particle within an interval $I$ arises in a natural way from the covariance structure itself.
Furthermore, this observable turns out to be compatible with the localization properties defined intrinsically as in 
\cite{BGL, FS02,Schroer97}.
In the last section we have also presented a concrete example, where this operator is compatible with the standard localization scheme. 

It is worth stressing that the proposed coordinate can be defined in every M\"obius-invariant theory, even if 
the group of symmetries is hidden, namely does not have a geometric action on the elements of the Hilbert space.
In those cases the physical meaning of the proposed coordinate operator and also of the intrinsic locality needs to be studied carefully.
This happens for example in free field theories on two-dimensional Minkowski spacetime, where a hidden $PSL(2,\bR)$ symmetry is present, and the usual locality properties differ from the intrinsic locality described above. 
Furthermore, the intrinsic locality turns out to be compatible with the locality shown by a dual theory defined on a null surface \cite{Sewell,MP}.
A similar situation arises in dual theories on null surfaces as  black hole horizon or null infinity of asymptotically flat spacetime, where the locality that arises on the null surface differs from the one present in the spacetime.

There are situations in which the modular dilations $\De^{it/(2\pi)}$ represent time translations, as for example in particular thermal theories. 
In those cases the  selfadjoint coordinate operator defined above, represents time. We stress that, since the corresponding energy $D$ is not bounded from below, there is no need to generalize the concept of observables by means of POVM as in the standard situation \cite{Naimark,Ludwig,POVM}. The procedure is very similar to the definition of a time operator in thermal theories \cite{NRT}, where, because of the unboundedness of the energy, 
a selfadjoint time operator can be defined.

Finally, we would like to emphasize once again the intrinsic character of the proposed coordinate operator, because its definition involves only the generators of the symmetry group.
This suggests that space and/or time could arise as a derived concept, namely as the target space of particular  observables representing spacetime coordinates.

\vsp
\noindent {\bf Acknowledgments}. 
The author is grateful to R.~Brunetti, S.~Doplicher, K.~Fredenhagen, V.~Moretti and M.~Porrmann for stimulating discussions and interesting suggestions. This work has been funded by Provincia Autonoma di Trento by the project FQLA, Rif. 2003-S116-00047 Reg. delib. n. 3479 \& allegato B, and partially supported by a grant from GNFM-INdAM (Istituto
Nazionale Di Alta Matematica) under the project {\it ``Olografia e spazitempo
asintoticamente piatti: un approccio rigoroso''}.

\appendix
\vspace{0.5cm}
\noindent
{\bf\Large Appendix}

\se{$PSL(2,\bR)$ representations\label{aa}}
The positive-energy irreducible representations $\pi_k$ of the covering group of $SL(2,\bR)$ are labelled by a real number $k\geq 1/2$ which coincides also with the lowest eigenvalue of the generator of rotations $(H+C)/2$ \cite{pukansky,sally}.
Every irreducible representation $\pi_k$ on the Hilbert space $L^2(\bR^+,dE)$ is generated by the following operators
\beq\label{gensl2r1}
H:=E,\qquad D=-i\sqrt{E}\frac{d}{dE}\sqrt{E},\qquad C=-\sqrt{E}\frac{d^2}{dE^2}\sqrt{E}+\frac{k^2-k}{E},
\eeq
which are defined on a suitable domain $\cD^k$ formed by linear combinations of the set of vectors 
$$
 {Z^{(k)}_m}(E) := \sqrt{\frac{\Ga(m-k+1)}{E\; \Ga(m+k)}} (2\be E)^k e^{-\be E} L^{(2k-1)}_{m-k}(2\be E)\;,
$$
where $L^{(k)}_n(x)$ are the generalized Laguerre polynomials and $\be$ is a positive constant.
$H,C,D$ satisfy the $sl(2,\bR)$ commutation relations:
$$
[H,D]=iH,\qquad [C,D]=-iC,\qquad [H,C]=2iD.
$$
The action of $H,C,D$ is closed on $D^k$. 
Moreover, $Z^{(k)}_m$ are analytic vectors for $H^2+C^2+D^2$.
Then by some result of Nelson \cite{nelson}, they are essentially selfadjoint, namely their selfadjoint extensions $\overline{H}$,$\overline{D}$,$\overline{C}$ are unique.
There, a representation of the covering group of $SL(2,\bR)$ on $\gH$ generated by 
$\overline{H}$,$\overline{D}$,$\overline{C}$ exists. 
In this paper we have discarded the bar and indicated by $H,D,C$ the selfadjoint operators.
\remark
{\bf (a)} In particular, if $k$ is integer, $\pi_k$ turns out to be a representation of $PSL(2,\bR)$ which is faithful if $k=1$. If instead $k$ is half-integer, $\pi_k$ is a representation of $SL(2,\bR)$ which is faithful for $k=1/2$.

\noindent
{\bf (b)} Since the vacuum is invariant under $\pi_k$, it can be extended to a representation on the Fock space.
Consider the complex conjugation $J$, its action on $H,D,C$ is as follows:
$$
JHJ=H, \qquad JDJ=-D, \qquad JCJ=C.
$$
The representation of $PSL(2,\bR)$ together with the above-defined $J$ is an irreducible representation of the M\"obius group.

\ssb{Another $sl(2,\bR)$-representation} On the one-particle Hilbert space $L^2(\bR^+,dE)$, consider the following operators
\beq\label{gensl2r2}
\widetilde{H}:=\frac{E^2}{2},\qquad \widetilde{D}=-i\frac{1}{2}\sqrt{E}\frac{d}{dE}\sqrt{E},\qquad \widetilde{C}=-\frac{d^2}{dE^2}+\frac{k^2-k}{E^2}.
\eeq
The set $\widetilde{\cD}^k$ formed by linear combination of the set of vectors 
$$
 {\widetilde{Z}^{(k)}_m}(E) := \sqrt{\frac{2\;\Ga(m-k+1)}{E\; \Ga(m+k)}} (2\be E^2)^k e^{-\be E^2} L^{(2k-1)}_{m-k}(2\be E^2)\;,
$$
turns out to be a dense set of common analytic vectors, thus they are essentially self-adjoint.
Furthermore, since they satisfy the $sl(2,\bR)$ commutation relations, they generate a positive-energy representation of the covering group of $SL(2,\bR)$ labelled by $k\geq 1/2$.
Here $k$ has the same meaning as the $k$ of the preceding sections (remark (a) above).

\end{document}